\documentclass[12pt]{revtex4}
\usepackage{amsmath,amssymb}

\begin{document}
\title{Brans-Dicke Gravity from Entropic Viewpoint}
\author{{\sc Ee Chang-Young}}
\email{cylee@sejong.ac.kr}
\author{{\sc Kyoungtae Kimm}}
\email{ktk@theory.sejong.ac.kr}
\author{{\sc Daeho Lee}\footnote{Present Address: 
\tt Center for Quantum Spacetime, Sogang University
Seoul 121-742, Korea}}
\email{dhlee@theory.sejong.ac.kr}
\affiliation{Department of Physics
and Institute of Fundamental Physics,
              \\Sejong University, Seoul 143-747, Korea}
\begin{abstract}
We interpret the Brans-Dicke gravity from entropic viewpoint.
We first apply the Verlinde's entropic formalism in the Einstein
frame, then  perform the conformal transformation which
connects the Einstein frame to the Jordan frame.
The transformed result yields  the equation of motion
of the Brans-Dicke theory in the Jordan frame.

\end{abstract}
\maketitle
\section{Introduction}
It is well-known that
a black hole has an entropy~\cite{beckenstein},
and a black hole radiates
as if it has
a temperature proportional to
the surface gravity on its event horizon~\cite{hawking_rad}.
These facts suggest that there may exist
a deep connection between gravity and thermodynamics.
Then, Jacobson~\cite{t_jacobson_95} showed that
the Einstein equation
can be obtained  from 
the first law of thermodynamics
together with the Bekenstein-Hawking entropy-area  relation~\cite{beckenstein,hawking_rad}
applied on a local Rindler horizon~\cite{Rindler_book}.

Recently, Verlinde~\cite{e_verlinde_01}
proposed that  gravity is
not a fundamental interaction
but can be explained  as a macroscopic
thermodynamic phenomenon.
Based on the holographic principle
and the equipartition rule,
he showed that gravity emerges
as an entropic force.
A related idea was considered also
by Padmanabhan~\cite{t.pad_10, t.pad_81}.
Since these works, there have appeared
many subsequent investigations
in cosmology~\cite{cai_cao_ohta},
black hole physics~\cite{EEKL1, EEKL2, TW:2010, LWW:2010},
loop quantum gravity~\cite{smolin_01},
and other fields.

In this paper we investigate whether the Brans-Dicke
gravity can be viewed as an entropic
phenomenon {\`a} la Verlinde.
The Brans-Dicke gravity is characterized by the fact that
the gravitational coupling is not presumed to be a constant but is
proportional to the inverse of a scalar field
which couples non-minimally to the curvature scalar~\cite{BD_theory}.
Also, it is well known that
Brans-Dicke theory can be transformed the Einstein frame
via a conformal transformation, and in the Einstein frame
Brans-Dicke action is equivalent to the Einstein gravity plus
the scalar field action except the fact that unlike in the
Jordan frame a usual matter does
not follow the geodesic of Einstein frame~\cite{BD_theory}.
We use the holographic principle and the equipartition rule
in the Einstein frame first, and then via a conformal transformation
we recover the field equations of metric in the Jordan frame.

In Ref.~\cite{G.Kang} it was shown that
black hole entropy in the Brans-Dicke gravity
depends upon the value of scalar field at
the horizon as well as the area of horizon.
However, we will show that a naive
application of Verlinde's formalism
with this entropy expression of the Jordan frame
cannot yield the correct field equations.

The paper is organized as follows.
First we  review the Verlinde's conjecture
on the emergent gravity
in the case of the Einstein gravity, then
give  an entropic derivation of
the Brans-Dicke field equation based on
the holographic principle and the equipartition rule.
We work with $c=\hbar=k_B =1 $ in this paper.

\section{Einstein Equations {\`a} la Verlinde}
According to Verlinde, gravity is an entropic force
emerging from coarse graining process of information
for a given energy distribution.
In this process, information is stored on the holographic screen.

In a static background
with a global timelike Killing vector $\xi^a$,
a generalized Newtonian potential is defined by
\begin{eqnarray}
\Phi={\log  }\sqrt{-\xi^a  \xi_a }.
\label{potential}
\end{eqnarray}
With this potential we can foliate the spacetime,
and we choose the holographic screen as
an equipotential surface of $\Phi$.
In this background, a test particle approaching
the holographic screen experiences
the following 4-acceleration
\begin{eqnarray}
\label{proper_acc}
a^a =
u^b \nabla_b    u^a
= - \nabla^{a}\Phi
\end{eqnarray}
where $u^a$ is the 4-velocity of the particle.
The temperature of the holographic screen
measured by an observer at infinity
is given by
the so-called  Unruh-Verlinde temperature
which is obtained
by multiplying the redshift factor
$e^{\Phi}=\sqrt{-\xi^{a} \xi_{a}}$
to the Unruh temperature for the acceleration (\ref{proper_acc}):
\begin{eqnarray}
T
= \frac{ 1}{2\pi}e^{\Phi} N^{a} \nabla_{a} \Phi,
\label{UV_temp}
\end{eqnarray}
where $N^a$ is the outward unit normal vector to the holgraphic
screen and the Killing vector.
With the holographic principle
which states that the information in the bulk
can be given by the information on the boundary,
and with the equipartition rule which states
that each bit of information contributes the energy of $\frac{1}{2}T$,
the quasi-local energy inside
the holographic screen which is the
boundary of spacelike hypersurface $\Sigma$ can be written as
\begin{eqnarray}
E = \oint _{\partial \Sigma} \frac{T}{2} dN
= \frac{1}{4\pi G} \oint _{\partial \Sigma}
e^{\Phi}N^{b} \nabla_{b} \Phi dA ,
\label{h_energy}
\end{eqnarray}
where the number of bits on an area $dA$ is assumed
to be $dN = dA/G$.
With (\ref{potential}) and the properties of Killing
vector $\xi^a$,
one  can show that the right hand side of (\ref{h_energy}) is
nothing but the Komar expression
\begin{eqnarray}
E_\text{Komar}=\frac{1}{4\pi G}
\oint _{\partial \Sigma}
dA_{ab} \nabla^{[a}\xi^{b]} ,
\label{komar_mass}
\end{eqnarray}
where $dA_{ab} =\frac{1}{2!} \epsilon_{abcd}dx^c \wedge dx^d$.
Expressing the Komar mass inside the holographic screen with
the energy momentum tensor $ T_{ab}$
in the left-hand side
and applying the Stokes theorem
to the right-hand side of Eq.~(\ref{komar_mass})
together with an identity, $\nabla^a \nabla_a \xi^b=-  R^{ab}\xi_a$
we have~\cite{R.M.Wald}
\begin{eqnarray}
2\int_{\Sigma}
\left(T_{ab }-\frac{1}{2}g_{a b } T^c_c\right)
 \xi^b  d\Sigma^a
= \frac{1}{4\pi G}
\int _{\Sigma}
R_{a b }\xi^b d\Sigma^a,
\label{einstein}
\end{eqnarray}
where $d\Sigma^a$ is defined with an outward pointing normal.
Since Eq.~(\ref{einstein}) is true for
an arbitrary holographic screen,
we obtain the Einstein equation
\begin{eqnarray}
R_{a b} =
8\pi G\left(T_{a b}
-\frac{1}{2}  g_{a b} T^{c}_{c}\right).
\end{eqnarray}


\section{Brans-Dicke Equations from Entropic Viewpoint}

In the previous section we have seen that Einstein gravity
can be seen as an entropic phenomenon.
In this section we will extend this entropic viewpoint of
gravity to the Brans-Dicke theory.
In the Brans-Dicke theory there are two frames called
the Jordan frame and the Einstein frame whose metrics
we denote by $\tilde{g}_{ab}$ and $g_{ab}$, respectively.
These two frames are related by
the following conformal
transformation:
\begin{equation}
\label{bd_conformal}
g_{ab}
= \psi \tilde{g}_{ab}.
\end{equation}
Unlike the minimal coupling in the Einstein frame,
the scalar field $\psi$ couples non-minimally to
the curvature scalar in the Jordan frame
such that it plays the role of effective gravitational
coupling of the theory, i.e., $G_\text{eff} = G/\psi$.
Since this extra scalar field comes into play, 
the gravity in the Jordan frame is also known as 
the scalar-tensor theory of gravity.
From now on, all quantities with tilde should be
understood as expressions in the Jordan frame.

In the following we will show that
the Brans-Dicke field equation of metric in the Jordan frame can
be obtained with  Verlinde's entropic formulation.
Since it is well known that the entropy of stationary black hole
in the Jordan frame is proved to be the same as that
in the Einstein frame~\cite{JKMPRD95},
first, we will setup entropic formulation in the Einstein frame
where the usual Verlinde's idea works nicely for the Einstein gravity
and then with a conformal transformation~(\ref{bd_conformal})
we will derive the field equations of
$\tilde{g}_{ab}$ in the Jordan frame.

As usual we assume  that
there exists a timelike killing vector $\tilde {\xi}^a$
in the  Jordan frame
satisfying the Killing equation,
$
\tilde{\nabla} _a \tilde {\xi}_b + \tilde{\nabla}_b \tilde{\xi}_a = 0
$
where $\tilde{\nabla}_{a}$ is the covariant derivative
associated with the Jordan metric $\tilde{g}_{ab}$.
Notice that $\xi_a = g_{ab}\tilde{\xi}^{b} $ satisfies the
Killing condition in the Einstein frame.

The gravitational potential $\Phi= \log\sqrt{ -\xi^a \xi_a}$
 associated with $\xi^a$ in the Einstein
frame has the following relation with the gravitational potential
$\tilde{\Phi}$ in the Jordan frame:
\begin{equation}
{\Phi}
=\tilde{\Phi} + \frac{1}{2}\log  {\psi},
\end{equation}
where
$\tilde{\Phi}=
\log \sqrt{-{\tilde{\xi}}^a {\tilde{\xi}}_a}$.
Therefore the acceleration of
a test particle near the holographic screen
seen in the Einstein frame is
given by
\begin{equation}
\label{einstein_acc}
{a}_a =- {\nabla}_a {\Phi}
=\tilde{a}_a - \frac{1}{2}\tilde{\nabla}_a \log {\psi},
\end{equation}
where the acceleration measured in the Jordan frame
is given by $\tilde{a}_a = - {\tilde{\nabla}}_a {\tilde{\Phi}}$.
Notice that unlike the Jordan frame description of the Brans-Dicke gravity, 
${a}_a = 0$ does not mean the geodesic motion of a test particle
in the Einstein frame. 
Eq.~(\ref{einstein_acc}) shows that a measure of
free-falling in the Einstein frame is given by $a_a +\frac{1}{2} \nabla_a\psi$
which compensates for the additional scalar field dragging.

With this acceleration $a_a$, we define the  Unruh-Verlinde temperature $T$ 
in the Einstein frame as follows:
\begin{equation}
\label{UV_temp_einstein}
{T} = e^{{\Phi}}{N}^a {a}_a
=  e^{\tilde{\Phi}}{\tilde{N}}^a
\left( \tilde{a}_a
+ \frac{1}{2} \tilde{\nabla}_a\log  {\psi}\right) ,
\end{equation}
where $N_a$ and $\tilde{N}_a$ are
the unit outward normal vectors to the holographic screen seen
in the Einstein frame and in the Jordan frame, respectively, and
are related by ${N}_a= \sqrt{ \psi } \tilde{N}_a$.
Therefore
the quasi-local energy (\ref{h_energy}) inside the holographic
screen $\partial \Sigma$ seen in the Einstein frame
has the following expression  
in terms of variables in the Jordan frame:
\begin{eqnarray}
{E}
= \oint_{\partial\Sigma} \frac{T}{2} \frac{dA}{G}
= \frac{1}{4\pi  G} \oint_{\partial  \tilde{\Sigma}=\partial  {\Sigma}}
e^{\tilde{\Phi}}{\tilde{N}}^a\left( \tilde{a}_a
 + \frac{1}{2} \tilde{\nabla}_a\log  {\psi}\right) \psi d\tilde{A},
 \label{bd_energy_E}
\end{eqnarray}
where the surface element is given by $dA= \psi d\tilde {A}$
under the conformal transformation~(\ref{bd_conformal}).
Here $ \partial \tilde{\Sigma}$ is the same closed hypersurface as 
 $ \partial {\Sigma}$ but seen in the Jordan frame. The same applies 
 to $ \tilde{\Sigma}$ and $\Sigma$ below.
Now, applying the Stokes theorem
to each terms in the right-hand side of Eq.~(\ref{bd_energy_E})
yields
\begin{eqnarray}
\frac{1}{4\pi G} \oint_{\partial  \tilde{\Sigma}}
e^{\tilde{\Phi}}{\tilde{N}}^a  \tilde{a}_a
 \psi d\tilde{A}
 =\frac{1}{4\pi G} \oint_{\partial  \tilde{\Sigma}}
 \psi \tilde{\nabla}^{[a}\tilde{\xi}^{b]} d\tilde{A}_{ab}
 =\frac{1}{4\pi G} \int_{\tilde{\Sigma}=\Sigma}
 \left(\psi\tilde{R}_{ab} -\tilde{\nabla}_a\tilde{\nabla}_b \psi
 \right) \tilde{\xi}^b d\tilde{\Sigma}^a ,
 \label{energy_g}
\end{eqnarray}
and
\begin{eqnarray}
\frac{1}{8\pi G}
 \oint_{\partial  \tilde{\Sigma}}
e^{\tilde{\Phi}}{\tilde{N}}^a
(\tilde{\nabla}_a\log  {\psi})\psi d\tilde{A}
=-\frac{1}{8\pi G}
 \oint_{\partial  \tilde{\Sigma}}
{\tilde\xi}^{[a}
 \tilde{\nabla}^{b]}\psi d\tilde{A}_{ab}
=-\frac{1}{8\pi G}
 \int_{ \tilde{\Sigma}={\Sigma} }
 \tilde{g}_{ab} \tilde{\square} \psi \tilde{\xi}^b d\tilde{\Sigma}^a,
 \label{energy_psi}
\end{eqnarray}
where $d\tilde{\Sigma}^a$ is the volume measure in the Jordan frame.
Thus, we have
\begin{equation}
\label{bd_energy2}
{E}
=\frac{1}{4\pi G} \int_{ \tilde{\Sigma}}
\left[ \psi \tilde{R}_{ab}
- \left(\tilde{\nabla}_a \tilde{\nabla}_b
+\frac{1}{2} \tilde {g}_{ab} \tilde{\square} \right)  \psi \right]
\tilde{\xi}^b d\tilde{\Sigma}^a,
\end{equation}
where $\tilde \square
= \tilde{g}^{ab}\tilde{\nabla}_a \tilde {\nabla}_b$.
As we have done in Eq.~(\ref{einstein})
we express $E$ in the terms of energy-momentum tensor as follows:
\begin{equation}
\label{bd_komar}
{E}
=2\int_{{\Sigma}}
\left({T}_{ab }
-\frac{1}{2}{g}_{a b }{T}_c^c \right)
 {\xi}^b     d{\Sigma}^a
= 2\int_{\tilde{\Sigma}=\Sigma }
\left(\tilde{T}_{ab }
-\frac{1}{2}\tilde{g}_{a b }\tilde{T}_c^c \right)
 \tilde{\xi}^b d\tilde{\Sigma}^a.
\end{equation}
Here, we have used the fact that the energy-momentum tensor
in the Einstein frame has the following relation with that
in the Jordan frame under
the conformal transformation (\ref{bd_conformal}),
\begin{equation}
T_{ab} = {\psi }^{-1} \tilde{T}_{ab}.
\end{equation}
We emphasize that the energy momentum tensor in (\ref{bd_komar}) 
contains the contribution from scalar field $\psi$ as well as
the ordinary matters.

Now, comparing two expressions (\ref{bd_energy2})
and (\ref{bd_komar}) for the holographic energy
we get the field equations for metric $\tilde{g}_{ab}$ as follows:
\begin{equation}
\tilde{R}_{ab}
= \frac{8\pi G}{\psi }
\left(
\tilde{T}_{ab} - \frac{1}{2} \tilde{g}_{ab} \tilde{T}_c^c  \right)
+\frac{1}{\psi}\left(\tilde{\nabla}_a \tilde{\nabla}_b
+\frac{1}{2}  \tilde{g}_{ab} \tilde{\square}
\right) \psi.
\label{bd_metric_eqs}
\end{equation}
This equation is the same one which can be obtained
by varying the Brans-Dicke action with respect to
the Jordan metric $\tilde{g}_{ab}$:
\begin{equation}
I_\text{BD} = \frac{1}{16\pi G}\int d^4 x   \sqrt{-\tilde{g}} \psi \tilde{R}
+ I[\tilde{g}_{ab}, \psi]
+ I_\text{matter}[\tilde{g}_{ab},\phi^\text{(m)}],
\label{bd_action}
\end{equation}
where $I[\tilde{g}_{ab}, \psi]$ is the action for scalar field $\psi$
and $I_\text{matter}[\tilde{g}_{ab}, \phi^\text{(m)}]$ is an
ordinary matter action minimally coupled to $\tilde{g}_{ab}$. 
Variation of last two terms in (\ref{bd_action}) with respect to $\tilde{g}_{ab}$
will contribute to
the energy momentum tensor of (\ref{bd_metric_eqs}).
Since the entropic consideration is restricted to the gravity sector, 
we cannot give explicit form of $I[\tilde{g}_{ab}, \psi]$ in (\ref{bd_action})
within our formulation.
However,  the coupling $\psi \tilde R$ in  (\ref{bd_action}) 
manifests the fact that the model we are considering is the Brans-Dicke gravity.
A prototype of  the scalar field action in the Brans-Dicke is usually given by
\begin{equation}
I[\tilde{g}_{ab}, \psi] =\frac{1}{16\pi G} \int  d^4x \sqrt{\tilde{g}} \frac{\omega}{\psi}
\tilde{g}^{ab} \tilde{\nabla}_a\psi  \tilde{\nabla}_b\psi ,
\end{equation}
where $\omega$ is the Brans-Dicke parameter. 

With the holographic principle and the equipartition rule
we have derived the field equations for metric of
the Brans-Dicke gravity without resort to the action principle
as Verlinde has done for the Einstein gravity.

Finally, we comment  that naive application of the holographic principle
and the equipartition rule in the Jordan frame does not yield
the correct field equations of Jordan metric.
Since in the Brans-Dicke gravity
 the entropy of black hole with horizon area $\tilde{A}$
 is given by~\cite{G.Kang}
\begin{equation}
\tilde{S}_\text{BH}
= \left. \frac{\psi \tilde{A}}{4G}\right|_\text{horizon},
\end{equation}
we expect that an infinitesimal area $d\tilde{A}$
on the holographic screen measured in the Jordan frame
would contain the following amount of information
\begin{equation}
d\tilde{N} = \frac{d\tilde{A}}{G_\text{eff}}
=\frac{\psi d\tilde{A}}{G}.
\end{equation}
Thus one may write the quasi-local energy
inside the holographic screen $\partial \tilde{\Sigma}$
of Jordan frame
which is compatible with the holographic principle
and the equipartition rule as follows
\begin{equation}
\tilde {E} =
\oint _{\partial\tilde{\Sigma}}
\frac{\tilde{T}}{2} \frac{\psi d\tilde{A}}{G}.
\label{jordan_energy}
\end{equation}
It is obvious from the Eq.~(\ref{bd_energy_E}) that
in order to obtain the Brans-Dicke field equations
the temperature $\tilde{T}$ in this expression
should be equal to the Unruh-Verlinde temperature
not of Jordan frame but
of Einstein frame in~(\ref{UV_temp_einstein}).




\begin{acknowledgments}
This work was supported by the National Research Foundation (NRF) of Korea grants
funded by the Korean government (MEST)
[R01-2008-000-21026-0 and NRF-2011-0025517].
\end{acknowledgments}

\end{document}